# ARAP: Demystifying Anti Runtime Analysis Code in Android Apps

Dewen Suo, Lei Xue, Runze Tan, Weihao Huang, Guozi Sun

*Abstract*—With the continuous growth in the usage of Android apps, ensuring their security has become critically important. An increasing number of malicious apps adopt anti-analysis techniques to evade security measures. Although some research has started to consider anti-runtime analysis (ARA), it is unfortunate that they have not systematically examined ARA techniques. Furthermore, the rapid evolution of ARA technology exacerbates the issue, leading to increasingly inaccurate analysis results. To effectively analyze Android apps, understanding their adopted ARA techniques is necessary. However, no systematic investigation has been conducted thus far.

In this paper, we conduct the first systematic study of the ARA implementations in a wide range of 117,171 Android apps (including both malicious and benign ones) collected between 2016 and 2023. Additionally, we propose a specific investigation tool named ARAP to assist this study by leveraging both static and dynamic analysis. According to the evaluation results, ARAP not only effectively identifies the ARA implementations in Android apps but also reveals many important findings. For instance, almost all apps have implemented at least one category of ARA technology (99.6% for benign apps and 97.0% for malicious apps).

## I. INTRODUCTION

Android, as one of the most popular mobile operating systems, has experienced exponential growth in its user base [1]. With millions of users worldwide, the Android platform has become a dominant force in the mobile industry. This popularity can be attributed to its open-source nature, extensive app ecosystem, and user-friendly interface. However, alongside its widespread adoption, Android apps face significant security challenges.

The Android platform has been a prime target for malicious actors due to its large user base and the potential for financial gain [2]. The proliferation of malware on the Android platform has become a serious concern. Malicious apps can compromise user privacy, steal sensitive information, and cause financial losses. In response to these threats, security researchers and analysts have been actively studying Android malware and developing effective countermeasures.

Meanwhile, to hide the malicious payloads, the malware developers are also starting to adopt anti-analysis techniques to hide them or raise the bar for analyzing them, including obfuscation [3]–[6] and app hardening [7]–[9] to evade detection and analysis. However, they have now shifted their focus towards anti runtime analysis (ARA) techniques.

The original purpose of ARA techniques was to protect intellectual property and prevent reverse engineering for legitimate purposes [8], [10]. However, their adoption by malware developers has complicated the efforts of security analysts [11], [12] to detect and analyze malicious behavior, thus prolonging the lifespan of their malicious apps.

To analyze Android malware, researchers have traditionally focused on three main categories of anti-analysis techniques[1]. The *first* category involves analyzing static anti-analysis techniques, primarily focusing on code deobfuscation. To identify the malicious code, security analysts need to deobfuscate the obfuscated code first. The *second* category involves code virtualization techniques [13] that convert the original code into a customized type of code. Therefore, security analysts need to first recover the semantics of the virtualized code (i.e., the customized type of code) and then determine whether it is malicious accordingly. The *third* category involves analyzing the code in hardened/packed apps [14]. Security analysts need to dynamically obtain the hidden actual implementations of the packed/obfuscated apps and then detect the malicious payloads statically.

However, there is still no systematic study on ARA techniques in Android apps, which hinders security analysts from effectively handling potential ARA implementations during malware analysis.

Previous works often either emphasize obfuscation techniques or focus only on certain aspects of ARA technology without conducting a systematic investigation of ARA technology and its concrete implementations and deployments [7], [15]–[18]. Moreover, the majority of these works involve only static analysis, which greatly limits their analytical capabilities.

Recently, dynamic analysis has been conducted on specific ARA techniques [19], [20]. Unfortunately, these studies do not offer a detailed investigation into ARA technology. They either superficially examine each type of ARA technique, focusing only on the most representative methods within each category, resulting in incomplete research, or they simply check if an app uses a certain ARA technique without delving into its specific implementation. This limits their usefulness to security analysts because only by understanding the specific implementation of ARA techniques can security personnel

---

Dewen Suo and Guozi Sun are with the School of Computer Science, Nanjing University of Posts and Telecommunications, No.9, Wenyuan Road, Yadong New District, Nanjing, 210023, China.

Lei Xue, Runze Tan, and Weihao Huang are with the School of Cyber Science and Technology, Sun Yat-sen University, No. 66, Gongchang Road, Guangming District, Shenzhen, Guangdong 518107, China.

Corresponding author: Guozi Sun (Email: sun@njupt.edu.cn).

[1]Please note the difference between anti-analysis techniques and ARA techniques. Anti-analysis techniques refer to a broader category that includes ARA and other techniques.

TABLE I
WORK CONTRAST. CONSIDER OUR WORK AS A UNIVERSAL SET (●), WHERE ○ REPRESENTS THAT WORK DOES NOT INVOLVE CERTAIN TECHNICAL ASPECTS, AND ◐ REPRESENTS PARTIAL SUMMARIES OF THOSE TECHNICAL ASPECTS.

|                      | AD | VED | AT | AH | RD |
|----------------------|----|-----|----|----|----|
| Evans et al. [15]    | ○  | ○   | ○  | ○  | ●  |
| Cho et al. [16]      | ◐  | ○   | ○  | ○  | ○  |
| Nguyen-Vu et al. [17]| ○  | ○   | ○  | ○  | ◐  |
| Shi et al. [18]      | ○  | ◐   | ○  | ○  | ○  |
| Sihag et al. [7]     | ◐  | ◐   | ◐  | ●  | ◐  |
| ATADetector [22]     | ◐  | ◐   | ●  | ●  | ○  |
| APPJITSU [19]        | ◐  | ◐   | ◐  | ◐  | ◐  |
| HALY [20]            | ◐  | ◐   | ◐  | ◐  | ●  |
| ARAP                 | ●  | ●   | ●  | ●  | ●  |

**AD**: Anti-Debugging  **VED**: Virtual Environment Detection
**AT**: Anti-Tampering  **AH**: Anti-Hooking  **RD**: Root Detection

take targeted measures. Furthermore, these dynamic works often rely on Frida [21], which, according to our investigation, is a prime target of ARA technology, presenting an apparent contradiction.

Furthermore, another significant issue is the lack of real-world application datasets, making it difficult to make accurate estimates regarding the implementation and deployment of ARA based on existing works. Table I provides a summary of the existing work, the terms of which are described in Section III-B.

In this paper, we aim to address this issue by conducting a systematic investigation of the ARA implementations adopted by off-the-shelf Android apps, including both benign and malicious ones. The investigation results and findings will assist analysts in gaining a better understanding of ARA technology and help them navigate around ARA implementations during malicious software analysis.

The purpose of this study is to provide developers and security analysts with a systematic view of ARA implementations in Android apps. By conducting a detailed empirical analysis, we aim to identify and evaluate the existing ARA implementations used in a diverse set of applications. Our ultimate goal is to assist analysts in bypassing these protections to effectively analyze malware. To achieve this, we performed a comprehensive empirical study focusing on the following research questions:

<u>Q1</u>: *How have ARA techniques evolved over time?*

<u>Q2</u>: *What are the differences between the ARA techniques adopted in various categories of apps?*

<u>Q3</u>: *What are the characteristics of ARA implementations in malicious apps?*

By answering Q1 and Q2, we aim to explore the normal evolution process and the current status of the ARA techniques, which can help developers and users know more about the security of their developed and used apps. Additionally, we want to assist the security analysts in analyzing the sophisticated malicious apps adopting ARA implementations, and hence we also answer Q3 to shed a light on how advanced malware evades analysis. In section IV-D, we will answer these questions one by one.

However, it is not straightforward to gain convincing answers to these questions, and a large-scale investigation is required with the following technical challenges.

*C1: The features of ARA techniques are not clear, and thus it is challenging to determine the ARA techniques adopted in the apps.* No specific studies have investigated the techniques adopted by Android apps to evade runtime analysis, and only limited rough summarizations of the potential ARA techniques are made [8], [10], [22]. Also, most of them focus on the techniques anti static analysis, such as code obfuscation [3], [4] and app packing [14], [23], [24]. There is still a lack of a comprehensive summarization of ARA techniques. The complexity of this issue lies in the continuous evolution and innovation of ARA techniques on the Android platform, with new techniques emerging constantly. Furthermore, the current research field faces a scarcity of available Ground Truth datasets, making it challenging to evaluate the effectiveness of experimental results.

*C2: Since the apps evade runtime analysis involving both static code and runtime behaviors, no existing analysis tools can be applied to investigating ARA implementations in apps.* To comprehensively analyze the runtime behavior of apps, it is crucial to develop an automated tool that can perform both static and dynamic analysis simultaneously. Manual completion of large-scale app analysis is impractical and time-consuming, highlighting the need for an automated approach to ensure efficient analysis. Additionally, the tool must effectively address the challenges posed by the evasion techniques employed by apps, including both static code and runtime behavior analysis.

For *C1*, in order to provide a comprehensive overview of ARA techniques, we have classified them into 5 major categories based on the standards defined by Mobile AppSec Verification Standard. To cover a wide range of techniques, we conducted a thorough review and synthesis of previous research work, documenting the typical characteristics of each category, including API call information, specific strings, etc. Building on this foundation, we performed semi-automated static analysis on real-world applications to identify the latest implementations of ARA techniques. This approach of reverse engineering actual application programs allows us to address real-world scenarios (§III-B). Furthermore, to evaluate the effectiveness of our tools, we created a Ground Truth dataset (§IV-C).

For *C2*, we developed an automated analysis tool called ARAP, which utilizes a hybrid approach combining static and dynamic analysis techniques. ARAP leverages call graphs to enhance the accuracy of static analysis and expands the coverage of elements in APK files through dynamic analysis techniques (§III). To circumvent various ARA techniques, we conduct dynamic analysis on real devices using a customized AOSP system. This system enables us to retrieve system API call information without triggering ARA mechanisms. By combining these methods, we are able to discover and understand ARA behaviors and implementations in Android apps in a more accurate and in-depth manner.

In summary, this paper makes the following major contributions:

TABLE II
MSTG Resilience requirement and corresponding ARA technologies

| MSTG Category | ARA technologies |
|---|---|
| MSTG-RESILIENCE-1 | Root Detection |
| MSTG-RESILIENCE-2 | Anti-Debugging |
| MSTG-RESILIENCE-3 | Anti-Tampering |
| MSTG-RESILIENCE-4 | Root Detection/Anti-Hooking |
| MSTG-RESILIENCE-5 | Virtual Environment Detection |
| MSTG-RESILIENCE-6 | Anti-Hooking |

- We conducted a comprehensive analysis by synthesizing previous research and performing semi-automated static analysis on real-world applications to conduct an in-depth analysis of ARA implementations, identifying 5 major categories and 32 subcategories. We further analyzed 117,171 APK samples collected from 2016 to 2023, including both benign and malicious APKs, to determine the deployment status and evolutionary trends of ARA technologies within them. Through this study, we have put forth a series of profound insights aimed at aiding security personnel in gaining a better understanding of application security and providing a foundation to counter the threats posed by malicious APKs.

- We have developed an automated detection tool called ARAP. This tool not only analyzes adversarial analysis behaviors but also examines other complex behaviors within apps. We plan to open-source this tool, making it a valuable resource for the research community and security professionals.

- To facilitate research in this field, we have constructed a ground truth dataset consisting of 300 APKs. As far as we know, this dataset is the first in the relevant research field and provides a valuable resource for future research and evaluation purposes.

By providing a comprehensive understanding of APK analysis within the context of Android ARA techniques, this research contributes to enhancing app security and detecting malicious behavior effectively. Security analysts can benefit from the insights and guidance provided, enabling them to analyze malicious apps more efficiently.

Our dataset, results, and source code of our framework are available at https://github.com/dfpp/ARAP.

## II. Background

### A. Android Anti Runtime Analysis Techniques

Android ARA techniques refer to a set of technical methods or approaches employed in Android applications to protect them from dynamic runtime analysis attacks. These attacks involve analyzing the behavior of an application during its execution to identify vulnerabilities or extract sensitive information.

The emergence of these ARA techniques is driven by the different motivations of benign application developers and malicious application developers. Benign application developers may use ARA techniques to protect their intellectual property, prevent reverse engineering, and deter malicious behavior. By employing ARA techniques, they can safeguard their code from being easily understood or tampered with by unauthorized individuals.

```java
private String[] virtualPkgs = {
    "com.bly.dkplat",
    "dkplugin.pke.nnp",
    // Package names of other known multiple app instances
};
public boolean checkByPrivateFilePath(Context context, VirtualCheckCallback callback) {
    // Get the private file path of the application.
    String path = context.getFilesDir().getPath();
    for (String virtualPkg : virtualPkgs) {
        // Checking for the existence of multiple app.
        if (path.contains(virtualPkg)) {
            // If the private file path contains a known package name of a multiple app.
            if (callback != null) {
                // Detected multiple app, invoking callback function
                callback.findSuspect();
            }
            return true; // Returning true indicates the existence of a known multiple app instance.
        }
    }
    return false;
    // Returning false indicates the absence of a known multiple app instance.
}
```

Fig. 1. A motivating example code

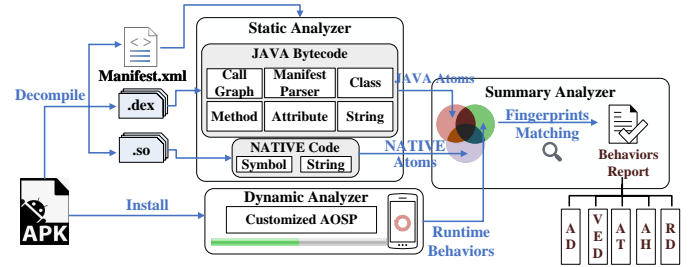

Fig. 2. Workflow of ARAP

On the other hand, malicious application developers exploit ARA techniques to conceal their malicious activities, evade detection and prevention mechanisms, and enhance the persistence of their applications. By employing ARA techniques, they make it more challenging for security researchers and analysts to understand the inner workings of their applications and identify any malicious behavior.

It is worth noting that although obfuscation is a common technique used to protect Android applications, it is not within the scope of this study, as it primarily hinders static analysis. Our focus is specifically on ARA techniques, which deal with protecting applications during runtime and thwarting dynamic analysis attacks.

To comprehensively classify ARA techniques, we follow the standards defined by Mobile AppSec Verification Standard (MASVS). MASVS sets a series of resiliency requirements for common tampering techniques, with the first six requirements focusing on the implementation of application defense techniques. In Table II, we provide a concise summary of the defense-related resilience requirements and their corresponding ARA classifications. A detailed description of MSTG_RESILIENCE can be obtained from https://github.com/OWASP/owasp-masvs/. We will provide a more detailed explanation of these techniques in Section III-B. However, due to space limitations, a complete description can be found in the Appendix.

## B. A Motivating Example

To better understand the ARA technology, we provide a code snippet in Figure 1 as a motivating example, which is taken from a real-world app.

In this code snippet, the app aims to check whether the current runtime environment is an app-level virtualized environment. Specifically, the code snippet achieves this goal by examining the File Path. The underlying principle is that the host application redirects the APK of the guest application to its own directory when loading it. In line 7, the app uses the *getPath* function to retrieve the path of the internal storage directory of the application and checks if the path contains common package names associated with multiple apps (lines 8-9). If a common package name is detected, the app invokes the callback function to perform the corresponding actions (lines 9-11). This detection method requires a pre-prepared list of known multiple app package names (lines 1-5). In Section III-A, we will introduce our detection method in detail.

## III. ANTI RUNTIME ANALYSIS PROBE

In this chapter, we will briefly describe the design architecture and working principle of ARAP (**A**nti **R**untime **A**nalysis **P**robe). Figure 2 illustrates the framework of ARAP, which can be divided into three parts: Static Analyzer (§III-C), Dynamic Analyzer (§III-D), and Summary Analyzer (§III-E). ARAP takes an Android APK file as input and attempts to output the reverse runtime analysis techniques deployed by the APK in the form of a JSON file. To circumvent various ARA techniques, particularly RD and AH, we perform dynamic analysis on real devices using a customized AOSP system. This system enables us to retrieve system API call information without triggering ARA mechanisms. It is worth mentioning that the static and dynamic analysis were performed independently. Thus, a failure in the dynamic analysis does not affect the static analysis, and vice versa.

### A. Definition of Atom and Fingerprint

Our detection is based on atoms and fingerprints, so it is necessary to introduce them first. We will illustrate our approach with a motivating example code.

In the motivating example code presented in Figure 1, we determine whether an app is running in a virtual environment by checking its private directory. This method iterates through the known package names in the virtual environment and compares them one by one with the app's private directory. If the app's private directory contains the package name of the virtual environment, then it is running in a virtual environment. This code uses the File Path method to detect App-Level Virtualization, and our goal is to detect whether the input APK has deployed this detection method.

We extract the key features of this code as fingerprints. For example, we select the package name of known multi-account apps as fingerprint A, which means that the code contains the package name of at least one known multi-account application. We select the functions *android.content.Context.getFilesDir()* and *android.os.File.getPath()* as fingerprint B, which implements private directory access. We select *java.lang.String.contains()* as fingerprint C, which indicates that the code performs string comparison. In addition, we need to ensure that fingerprint B and fingerprint C appear in the same code segment. We introduce the call graph (CG) to address this issue. In the call graph, fingerprint B and fingerprint C are considered valid only if they appear in the same code segment. It is noted that fingerprint B consists of two methods, and we define *android.content.Context.getFilesDir()* and *android.os.File.getPath()* as atom A and atom B respectively. Fingerprint B is valid only when both atoms A and B appear simultaneously. In this way, we define the fingerprint of this code segment. When fingerprint A, fingerprint B, and fingerprint C appear simultaneously, we consider that the APK has deployed the Anti App-Level Virtualization technique, specifically, by using the File Path method. In order to accurately identify the File Path method implemented in other forms, we conducted a semi-automated static analysis to establish a comprehensive list of atoms and fingerprints for this method (§III-B). The complete atom and fingerprint list of all ARA techniques will be provided along with the ARAP tool.

To ensure the effective operation of ARAP, we analyzed and investigated a substantial number of real-world implementations of ARA technology. Ultimately, we identified 1,515 distinct atoms and 175 unique fingerprints to recognize all ARA technologies mentioned in Section III-B. The results in Section IV-C demonstrate that while ARAP utilizes the relatively conventional approach of atoms and fingerprints, this method proves to be both simple and effective.

In the spirit of transparency and collaboration, we have chosen to open-source ARAP. The tool is written in Python and encompasses more than 7,600 lines of code (excluding the code of Androguard itself).

### B. Identifying ARA Techniques

To encompass a broader range of technical details, we first conducted a comprehensive review of previous research work and documented the typical characteristics of each category of techniques. This includes API call information, specific strings, such as "*isDebuggerConnected()*" and "re.frida.server". Subsequently, we leveraged these typical characteristics to perform semi-automated static analysis on real-world apps. In other words, we manually explored all possible ARA techniques using these typical characteristics as entry points. This allowed us to discover the latest ARA techniques employed in the real world. Ultimately, we identified 32 subcategories of ARA techniques, with multiple implementation variations for each subcategory. The implementation details of various ARA techniques can be obtained from the source code of the tool we provide.

*1) Anti-Debugging (AD):* Anti-Debugging is a technique used to protect applications from reverse engineering [25]. This technique involves inserting anti-debugging code into the application code, making it difficult for analysts to use debuggers and additional tools to debug and reverse engineer the application. Deploying appropriate anti-debugging techniques can effectively enhance the security and resilience of applications against malicious attacks [26]. Table III provides a summary of the AD techniques.

TABLE III
ANTI-DEBUGGING (AD) IMPLEMENTATIONS

| AD Technology | Principle/Detail |
|---|---|
| Using Timing Check (UTC) | Time Differences |
| Using Signals (US) | Signal Processing Differences |
| Using TracerPID (UTP) | Values of /proc/PID/status |
| Preempt Ptrace API (PPA) | Preemptively Attaching to Self |
| Anti-Dumping (ADU) | Utilizing Inotify Events |
| Debuggable and Debugger Features (DDF) | Leveraging System APIs and Debugger Features |
| Using Breakpointset Structure (UBS) | Detecting Breakpointset Structure |
| Altering Debugger Memory (ADM) | Modifying the Global Virtual Machine State |

TABLE IV
EMULATOR DETECTION (ED) IMPLEMENTATIONS

| ED Technology | Principle/Detail |
|---|---|
| Device Identifier (DI) | Checking for Specific Values or Nullness |
| Current Build (CB) | Checking for Specific Values |
| Telephony Manager (TM) | Validity of TM Service |
| Hardware Composition (HC) | Validity of Hardware Components |
| Emulator Related Strings (ERS) | Checking for Specific Strings Related to Emulator |
| Sensor Information (SI) | Validity of Sensor Information |
| Context Switch Based Detection (CSBD) | No Race Conditions in the Same Basic Block of QEMU |
| Translation Block Cache Based Detection (TBCBD) | Undermining QEMU's Translation Block Cache |
| Misaligned Vectorization Based Detection (MVBD) | Utilizing Misaligned Vectorization |
| Updates to Execution State Bits (UESB) | Actively Flipping the E-bit of CPSR |

TABLE V
APP-LEVEL VIRTUALIZATION DETECTION (ALVD) IMPLEMENTATIONS

| ALVD Technology | Principle/Detail |
|---|---|
| File Path (FP) | Checking the Current App's Path |
| Share UID (SU) | Existence of Identical UID |
| Port Detection (PD) | Leveraging Port Communication |
| Code Injection and Hooking (CIH) | Checking Stack and SO Files |

TABLE VI
ANTI-TAMPERING (AT) IMPLEMENTATIONS

| AT Technology | Principle/Detail |
|---|---|
| Signature Checking (SC) | Relying on Digital Certificates |
| Code Integrity Checking (CIC) | Checking Integrity of Specific Code |
| SafetyNet Attestation (SA) | Utilizing SafetyNet Attestation API[†] |
| Installer Verification (IV) | Checking the Installation Source |

[†] The Play Integrity API is also included in our detection scope.

*2) Virtual Environment Detection (VED):* The main purpose of this technique is to detect whether a device is running in a virtual environment [27], [28]. For malicious applications, by detecting whether the device is running in a virtual environment, they can decide whether to exhibit malicious behavior to evade detection [29], [30]. For benign applications, developers determine whether the application is running in a secure environment [10], [31], [32].

*Emulator Detection (ED).* Emulator Detection is a type of software or technology used to detect and defend against applications running in emulators or virtual machine environments against malicious attacks, anti-debugging, cracking, or other fraudulent behavior. Table IV presents the specific techniques of ED.

*App-Level Virtualization Detection (ALVD).* Android App-Level Virtualization technology is an innovative virtualization technique that allows loading and running any client application's APK file within a host application [18], [33]. Some traditional emulator detection methods may fail to detect it [18], [34]. Table V displays the specific methods of ALVD technology.

*3) Anti-Tampering (AT):* "Anti-Tampering" is designed to prevent malicious attackers from tampering or repackaging applications. Implementing an Anti-Tampering mechanism in an application can effectively protect it from malicious modifications, thereby ensuring the security and integrity of the application [22], [35], [36]. Table VI provides a summary of the AT techniques.

*4) Anti-Hooking (AH):* Android Hooking techniques refer to the use of hooking technology by attackers to monitor, modify, or tamper with the behavior of an application, thereby carrying out malicious attacks or stealing sensitive information [8], [37]. Commonly used tools include Xposed and Frida.

*Xposed.* Xposed is an open-source hooking framework that allows developers to modify the behavior of applications by writing modules.

*Frida.* Frida is a dynamic instrumentation tool that allows for hooking into applications using JavaScript scripts. It can be used for monitoring and modifying aspects of an application such as memory, files, and network activity.

*5) Root Detection (RD):* In Android, Root refers to the process of gaining superuser privileges. With Root, users can access all the features and files of the system, and thus can modify system settings, delete pre-installed applications, install unverified applications, and so on [17]. The details of the RD techniques can be found in Table VII.

## C. Static Analyzer

The static analysis part of ARAP can be further divided into two subsections: static analysis of the JAVA Bytecode and static analysis of the NATIVE Code.

For the analysis of the JAVA Bytecode, we utilized the well-known Android analysis tool Androguard [38]. However, the default automated analysis template provided by Androguard can only handle single-dex APKs. Therefore, we firstly extract all the dex files by decompressing the APK so that Androguard can parse the entire APK file completely. For JAVA Bytecode, we extract classes, methods, strings, and attributes and compare them with our known atoms. Additionally, as mentioned in the example code earlier, we introduced call graphs to reduce false positives. In addition, we also parse the manifest.xml file. This is because the manifest.xml file can define whether the APK file can be debugged, and the IV technology can be enabled through simple settings in the manifest.xml file.

6TABLE VII
ROOT DETECTION (RD) IMPLEMENTATIONS

| RD Technology | Principle/Detail |
|---|---|
| Check Installed Packages (CIP) | Checking for the Installation of Root-Related Apps |
| Check Shell Commands Execution (CSCE) | Checking for the Execution of Privileged Commands |
| Check Build Tag and System Properties (CBTSP) | Leveraging Build Tags and System Properties |
| Directory Permissions (DP) | Checking Permissions of Specific Directory |

For NATIVE Code, we use the nm and strings commands to extract symbols and strings from each .so file, respectively. Additionally, we analyze the binary form of each .so file to determine whether it uses specific machine code.

The Static Analyzer assists in efficiently pinpointing the traces of ARA technology deployment in developers' applications. To expedite the analysis process and alleviate the burden on the Dynamic Analyzer, the utilization of the Static Analyzer becomes essential. It gathers all identified atoms and forwards them to the Summary Analyzer, enabling the latter to determine the specific ARA techniques employed in the app based on a predefined list of atoms and fingerprints.

### D. Dynamic Analyzer

As mentioned earlier, ARA technologies are complex, and static analysis alone cannot achieve recognition of all ARA technologies. This means that dynamic analysis is necessary. In order to accelerate the analysis speed and minimize the triggering of time-related ARA techniques (UTC), we selectively gather essential API calls, focusing on behaviors that cannot be accurately determined through static analysis alone.

We use "adb install -g" command to install the APK, and then use the Monkey tool to randomly generate 10,000 events to explore the app, in order to trigger as many different functions of the app as possible. Finally, we restart the app and repeat the above process three times. Throughout this process, we use a customized AOSP to capture the APIs invoked by the app, in order to determine if it employs any ARA techniques. As shown in Figure 2, the Dynamic Analyzer captures "Runtime Behaviors" and provides them as atoms input to the Summary Analyzer. For example, Dynamic Analyzer would focus on the native layer's *ptrace* function to determine whether the APP actively invokes it to employ PPA techniques in order to prevent debugger attachment. In this example, the Dynamic Analyzer is necessary as we need to ascertain the parameters used when calling the *ptrace* function, which cannot be obtained by Static Analyzer.

### E. Summary Analyzer

The detection results from both the Static and Dynamic Analyzers are fed into the Summary Analyzer, which integrates the atoms reports generated by both Analyzers. For each APK, the Summary Analyzer generates two reports based on predefined fingerprints—a simplified version and a detailed version.

The simplified version of the report classifies the ARA techniques employed by the APK into 5 major categories. This allows security analysts to swiftly grasp the technology types utilized by the APK, facilitating an initial assessment and classification. In contrast, the detailed version of the report offers a more comprehensive and meticulous breakdown, providing insights into 32 distinct subcategories. Through this detailed report, security analysts can gain a thorough understanding of the specific technical intricacies employed by the APK. This aids in accurately comprehending the APK's behavior and enables the implementation of corresponding countermeasures.

## IV. EVALUATION AND EMPIRICAL STUDY

In this section, we will first introduce our experimental environment and the dataset used. Following this, we present the results of our empirical study on the deployment of ARA techniques for different types of apps. Our study focuses on both popular and regular benign apps, as well as malicious apps. We analyze the characteristics and temporal trends of these apps, comparing the differences between benign and malicious APKs. This study offers insights into the current state of ARA techniques deployment and identifies potential areas for improvement in this field.

### A. Experimental Environment

Our study employed four desktop computers equipped with Intel Xeon E-2224G processors and 16GB of memory. These computers were running the Linux-based operating system CentOS Linux 7 (Core), with Python version 3.8.15 installed. The test phones used in the study were Pixel 4, with the AOSP version android-13.0.0_r31.

### B. Dataset

Our work involves 117,171 APKs, which we divided into four datasets. Our experiments will be conducted based on these datasets.

**PopularBench.** In the popular free ranking list of Google Play Store, there are a total of 34 categories. We downloaded the top 50 most popular apps in each category, resulting in a total of 1650 APKs, which compose the PopularBench dataset[2].

**BenignBench.** We obtained a total of 80,469 benign APKs from AndroZoo, spanning from 2016 to 2023. We consider an APK to be benign only if the VirusTotal field is 0. These APKs constitute the BenignBench dataset. The specific number of APKs acquired for each year is shown in Table VIII.

**MaliciousBench.** Similar to the BenignBench dataset, we obtained a total of 24,716 malicious APKs from AndroZoo, spanning from 2016 to 2023. We consider an APK to be malicious only if the VirusTotal field is greater than 10. These APKs constitute the MaliciousBench dataset, and the specific quantity details are provided in Table VIII.

TABLE VIII
DISTRIBUTION OF APKS IN THE BENIGNBENCH AND MALICIOUSBENCH DATASETS

| | 2016 | 2017 | 2018 | 2019 | 2020 | 2021 | 2022 | 2023 | Total |
|---|---|---|---|---|---|---|---|---|---|
| BenignBench | 9,969 | 10,000 | 10,000 | 9,864 | 10,000 | 10,000 | 20,636 | 11,918 | 92,387 |
| MaliciousBench[†] | 4,872 | 4,999 | 4,969 | 4,583 | 2,021 | 3,044 | 228 | 68 | 24,784 |

[†] The number of malicious APKs in specific years from AndroZoo is relatively low, and we have collected all of them.

[2]The Watch faces category only had three APKs, so we excluded it.

**GroundTruth.** GroundTruth consists of 300 different APKs collected between 2016 and 2023. These APKs were randomly obtained from AndroZoo and are used to measure the performance of ARAP.

*C. Performance of ARAP*

To ensure the credibility of our experimental results, we utilize the GroundTruth dataset to measure the performance of ARAP. When establishing the foundational dataset, three members of our team independently analyzed these APKs and annotated the employed ARA techniques. The ground truth primarily involves static analysis and dynamic analysis. During static analysis, the annotators first attempt to analyze the program statically and quickly locate relevant code segments using keywords. They then analyze whether the identified code segments involve ARA techniques and further confirm the specific type of ARA technique, if applicable. During dynamic analysis, the annotators run the APK on a "clean" real device without any debugging behavior and observe and record its behavior as a baseline. Then, the annotators sequentially run the APK in different environments, such as emulators and rooted devices, and attempt to debug the APK using tools like Frida to observe any behavioral changes. If the behavior differs from the baseline, further investigation is conducted to determine if it is caused by ARA techniques. Finally, the annotators combine the results from static and dynamic analysis to annotate the APK. Only when all three annotators reach a consensus on the same APK, the corresponding annotation results are included in the ground truth.

Due to the complexity of ARA techniques, manual analysis requires a significant amount of time. To ensure reliability, each application required 3 to 8 hours of analysis. Despite the time constraints, our Ground Truth dataset still consists of 300 APKs. As far as we know, this is currently the first available dataset in the relevant field.

For the 300 APKs, there were a total of 1500 major categories to be detected and 9600 subcategories to be detected[3]. ARAP accurately detected all ARA techniques in the major categories without any false positives or false negatives, achieving a precision and recall of 100%. In the case of subcategories, ARAP had a total of 7 false positives but no false negatives, resulting in a precision of 99.6% and a recall of 100%. Among the 7 false positives, 2 originated from the native layer of the APKs. Due to ARAP's simple binary code comparison, unrelated bytecode was mistakenly identified as a specific ARA technique. The remaining 5 false positives were from the Device Identifier section of the emulator detection. As some APKs only relied on a simple check of whether the Device Identifier was null to determine if the current environment was an emulator, ARAP couldn't extract sufficient valid information, resulting in false positives. It is important to note that even though there were false positives in the subcategory detection, it did not impact the performance in detecting major categories.

Furthermore, we conducted a comparison with ATAdetector [22]. We had ATAdetector analyze the 300 APKs in the Ground Truth dataset and reclassify its detection results to align with the 10 subcategories in the ARAP detection items. In the end, the ATAdetector successfully analyzed a total of 292 APK files. It is worth noting that our work overlaps with theirs in 10 subcategories, while the remaining 22 subcategories are unique additions from our research. It is important to note that the APKs in the Ground Truth dataset are not the same as those analyzed during the initial collection of atoms and fingerprint lists. For the most relevant works, [7] did not utilize any tools, [19] used tools that are not open source, and [20] provided the open source URL of the tools, but unfortunately, the website is currently inaccessible.

[3]Each APK requires the detection of 5 major categories and 32 subcategories.

TABLE IX
PERFORMANCE COMPARISON OF ATADETECTOR AND ARAP

|  | Accuracy | F1 | Precision | Recall |
| --- | --- | --- | --- | --- |
| ATAdetector | 0.868 | 0.792 | 1 | 0.657 |
| ARAP | 0.997 | 0.998 | 0.996 | 1 |

Table IX presents a performance comparison between ARAP and ATAdetector. ARAP exhibits outstanding performance across all four metrics, particularly excelling in Accuracy, F1, and Recall when compared to ATAdetector. With a Recall score of only 0.657, ATAdetector appears to have missed a significant number of ARA technologies. These results indicate the high accuracy and reliability of ARAP in correctly identifying the ARA technology.

The exceptional performance of ARAP is attributed to our extensive analysis and investigation of numerous real-world applications, which allowed us to identify a significant number of valid atoms. In contrast, ATAdetector's approach to detecting ARA technology primarily refers to the official Android documentation and OWASP security guidelines, lacking the exploration of real-world ARA implementations. Consequently, it struggles to effectively identify ARA technology in real-world applications.

*D. Research Questions*

To investigate the deployment of ARA techniques for different types of apps, we set out to answer the following research questions:

**Q1: How have ARA techniques evolved over time?**
**Q2: What are the differences between the ARA techniques adopted in various categories of apps?**
**Q3: What are the characteristics of ARA implementations in malicious apps?**

Through answering these research questions, we aim to provide a comprehensive understanding of the current state of ARA techniques deployment and identify potential areas for improvement.

💡 **Q1: How have ARA techniques evolved over time?**

In Q1, we investigated the deployment and usage of ARA techniques in benign apps, aiming to understand the current state and evolving trends in app security. Through this investigation, we gain insights into the security landscape of apps and their ability to withstand malicious attacks. Additionally, we observe a trend in the adoption of ARA techniques within apps. These findings are of significant importance in assessing

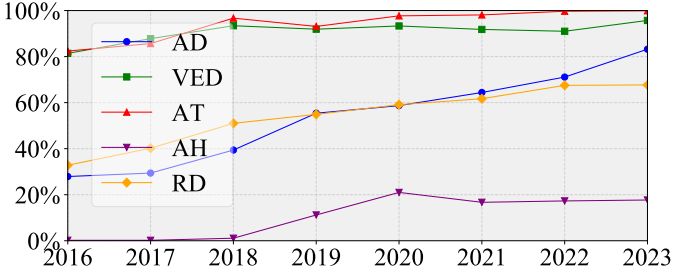

Fig. 3. Evolution of ARA Techniques in Benign Apps Deployment over Time

the security awareness of developers and enhancing the overall security performance of apps.

We conducted an initial investigation into the overall evolution of various ARA technologies between 2016 and 2023, aiming to assess developers' level of security awareness. Figure 3 depicts the results of our survey. From the line graph, it is evident that all five types of technologies show an overall upward trend. Among them, VED technology has remained relatively stable over the past six years, while AH technology has maintained a stable state in the past four years.

Among these categories, the AD technology exhibits a significant increase from 27.9% in 2016 to 83.2% in 2023. The VED technology demonstrates a relatively stable deployment rate, ranging from 81.4% in 2016 to 95.7% in 2023. The AT technology shows a gradual increase, reaching 99.3% in 2023 from 82.4% in 2016. AT and VED technologies stand out with significantly higher deployment rates compared to others, placing them in the top tier. On the other hand, the AH technology receives minimal attention, with a deployment rate below 20% throughout the years. It starts at 0.2% in 2016 and increases to 17.7% in 2023. RD technology shows a consistent upward trend, rising from 32.8% in 2016 to 67.7% in 2023.

Overall, the data indicates a positive trend in the adoption of security measures by developers, with varying degrees of attention given to different technologies. However, there is still room for improvement, particularly in enhancing the focus on AH technology.

Table X to XII provide a detailed overview of the deployment of various technologies in benign apps in 2023. In terms of AD technology, DDF stands out with a usage rate of 83.0%, leaving UTP and US in the second and third positions with usage proportions of only 9.1% and 5.9%, respectively. The remaining AD technologies have usage proportions below 5%. When it comes to VED technology, CB, DI, and ERS rank 1st, 2nd, and 3rd respectively, with proportions of 87.2%, 69.1%, and 41.1%. Compared to AD technology, VED exhibits a greater diversity in the techniques used, with 5 technologies having usage proportions above 10%. However, it is worth noting that the majority of deployed VED technologies are targeted towards traditional emulators, with a ratio of approximately 12:1. This also indicates that developers are not adequately prepared to address the newer app-level virtualization technologies.

For AT technology, the deployment proportion of SC is an astonishing 97.3%. This signifies that almost all developers possess the fundamental awareness of app protection, which is a pleasantly surprising data. It is worth noting that the signing mechanism of Android V2 and later versions has a strong anti-tampering capability. We consider an app to have deployed SC technology only if it uses V2 or later versions of signing or performs runtime dynamic signature verification. Furthermore, IV and CIC also exhibit high proportions, reaching 77.3% and 66.5% respectively. As mentioned earlier, AH technology is the least prioritized among developers. In this regard, Xposed accounts for a proportion of 17.2%, while Frida only reaches 3.6%. This indicates that the majority of developers have not taken into consideration the possibility of their apps being analyzed using Frida. This is a matter that deserves the attention of developers, who should utilize certain AH techniques to safeguard their applications from potential attacks. In the realm of RD technology, the deployment proportions of CBTSP, CSCE, and CIP are relatively equal, at 57.8%, 51.6%, and 50.2%, respectively. However, DP has a comparatively lower proportion, accounting for only 11.2%.

TABLE X
DETAILED ANALYSIS OF BENIGN APPS IN TERMS OF AD

| AD Technology | DDF | UTP | US | ADM | UBS | ADU | PPA | UTC |
|---|---|---|---|---|---|---|---|---|
| Quantity | 9,892 | 1,083 | 698 | 264 | 156 | 16 | 13 | 10 |
| Percentage | 83.0% | 9.1% | 5.9% | 2.2% | 1.3% | 0.1% | 0.1% | 0.1% |

It is essential for software developers to utilize ARA technology in their applications. It helps safeguard the security of user data, prevents malicious attacks on applications, and enhances the reliability and stability of the applications.

> **Q1 Finding**
>
> Over the years, there has been a progressive increase in the security awareness among Android developers. Simultaneously, we observe that simpler and more efficient techniques are favored by developers and thus have higher adoption rates. Finally, we find that there is still room for improvement in terms of developer attention to AH technology.

💡 **Q2: What are the differences between the ARA techniques adopted in various categories of apps?**

In Q2, we will focus on studying whether there are significant differences in the deployment and usage of ARA techniques between popular apps and regular apps. Popular apps are commonly expected to be more secure and reliable. Additionally, we also aim to investigate whether this deployment pattern is correlated with the app categories. We employed the apps from BenignBench as a representation of regular apps. To avoid any potential interference caused by multiple versions of the same app, we only retained the most recent version of each app.

Figure 4 depicts a comparison between popular apps and regular apps in terms of their deployment of ARA techniques in the year 2023. In the graph, we can observe that popular apps indeed outperform regular apps in the utilization of AD, VED, AT, and RD technologies. Particularly, in terms of RD deployment ratios, popular applications are significantly ahead with a leading percentage of 25.1%, indicating that these applications place a higher emphasis on virtual environment detection. Additionally, all popular apps have implemented AT technology, which is both surprising and reasonable. The only exception arises in the case of AH, where popular apps seem to prioritize it less, with a proportion of only 9.2%. We once again emphasize the importance of deploying AH technology, and developers should further enhance their security awareness to incorporate AH techniques into their apps.

TABLE XI
DETAILED ANALYSIS OF BENIGN APPS IN TERMS OF VED

| VED | ED | | | | | | | | | | ALVD | | | |
|---|---|---|---|---|---|---|---|---|---|---|---|---|---|---|
| | CB | DI | ERS | SI | HC | MVBD | TM | UESB | CSBD | TBCBD | SU | FP | CIH | PD |
| Quantity | 10,388 | 8,231 | 4,893 | 2,053 | 1,465 | 561 | 327 | 13 | 9 | 0 | 835 | 234 | 120 | 18 |
| Percentage | 87.2% | 69.1% | 41.1% | 17.2% | 12.3% | 4.7% | 2.7% | 0.1% | 0.1% | 0% | 7.0% | 2.0% | 0.7% | 0.2% |

TABLE XII
DETAILED ANALYSIS OF BENIGN APPS IN TERMS OF AT/AH/RD

| AT/AH/RD | AT | | | | AH | | RD | | | |
|---|---|---|---|---|---|---|---|---|---|---|
| | SC | IV | CIC | SA | Xposed | Frida | CBTSP | CSCE | CIP | DP |
| Quantity | 11,597 | 9,214 | 7,931 | 945 | 2,050 | 430 | 6,889 | 6,150 | 5,983 | 1,335 |
| Percentage | 97.3% | 77.3% | 66.5% | 8.0% | 17.2% | 3.6% | 57.8% | 51.6% | 50.2% | 11.2% |

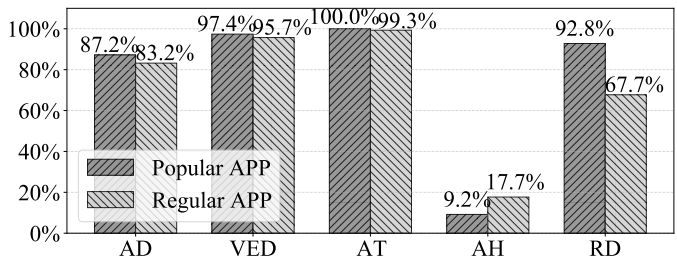

Fig. 4. Comparison between Popular Apps and Regular Apps

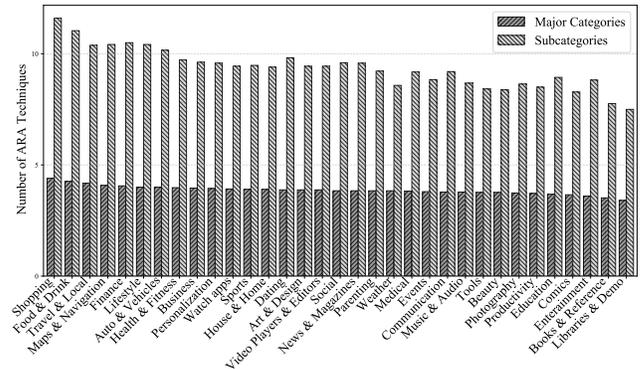

Fig. 5. Average number of ARA techniques implemented in an app per category.

In addition, we would like to further understand the deployment status of ARA technologies in relation to different categories of apps. Generally, people expect ARA technologies to be more prevalent in certain categories of apps. We have conducted a survey to determine the average number of ARA technologies implemented in each category of apps. Figure 5 presents the survey results, where the labels on the x-axis are arranged in descending order based on the values of the Major Categories.

Overall, the top three categories that utilize ARA technology the most are Shopping, Food & Drink, and Travel & Local. The performance of the Food & Drink category is particularly surprising, as traditionally, apps in this category may not be considered to have high security requirements. Moreover, the Shopping, Finance, and Business categories, which are traditionally perceived to have higher security demands, rank high in the number of ARA implementations, aligning with expectations. Entertainment, Books & Reference, and Libraries & Demo are the three categories with the least utilization of ARA technology. This is consistent with the conventional understanding that these categories may not require a high level of security. Therefore, developers may prioritize enhancing app functionality for performance considerations.

It is worth noting that even in the Libraries & Demo category, which has the lowest usage of ARA technology, an average of 3.4 ARA major categories and 7.5 ARA subcategories are employed per app. In our future work, we aim to evaluate the effectiveness of different categories of ARA technology to assist developers in achieving higher security with minimal additional costs.

> **Q2 Finding**
>
> In general, popular apps tend to have higher deployment proportions of ARA technologies compared to regular apps, with AH technology being an exception where popular apps perform less favorably than regular apps. In terms of categories, it is consistent with the general expectation that categories traditionally perceived to have higher security requirements utilize more ARA technology.

### 💡 Q3: What are the characteristics of ARA implementations in malicious apps?

In Q3, our objective is to explore the ability of malicious apps to resist security analysis. We will compare malicious apps with benign apps and identify and summarize the unique characteristics of malicious apps in deploying ARA techniques. At the same time, we are aware of the increasing difficulty of detecting malicious apps. Therefore, we hope to provide new ideas and methods for malicious app detection through our research.

We first examined the progression of deploying different ARA techniques in malicious apps over time. As depicted in Figure 6, there is an overall upward trend in the deployment of AD, AT, AH and RD technologies in malicious apps. This indicates that the analysis and identification of malicious apps is becoming more challenging. However, VED stands out as an exception, reaching its peak in 2017 at 64.9%. It experienced fluctuations between 2018 and 2021 but showed a recovery in 2023, reaching 53.0%.

Figure 7 presents a more intuitive comparison of the deployment of the five types of techniques in benign and malicious apps over the past four years. The most significant difference lies in VED, where the proportion of malicious apps deploying VED is only 33.0%, much lower than the 92.6% observed in benign apps. This finding is unexpected, as only a small portion of malicious apps attempt to detect the current runtime environment to determine whether to exhibit malicious behavior. On the contrary, benign applications are more inclined to assess the current runtime environment. Furthermore, for AD, AT and AH technologies, the deployment proportions in benign apps are slightly higher than those in malicious apps. This indicates that benign apps perform better in terms of "protecting the app". However, the deployment proportion of AD, AT, and RD technologies in malicious apps still remains above 50%. This indirectly indicates that analyzing and identifying malicious apps have become increasingly challenging. The only exception arises in the case of RD technology, where malicious apps tend to prefer devices that have been rooted. This is because a rooted device grants malicious apps greater privileges to execute malicious activities.

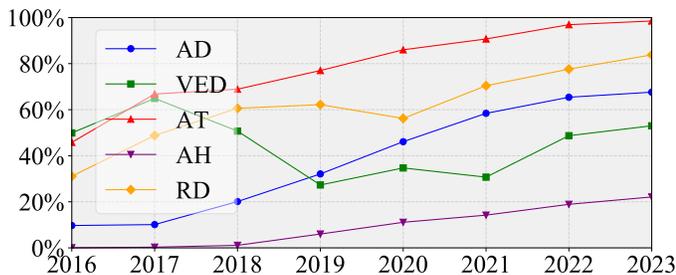

Fig. 6. Evolution of ARA Techniques in Malicious Apps Deployment over Time

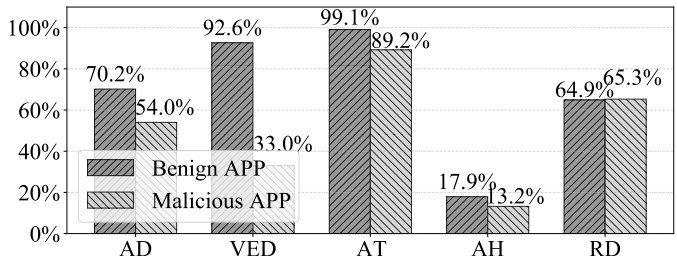

Fig. 7. Comparison between Benign Apps and Malicious Apps

TABLE XIII
DETAILED ANALYSIS OF MALICIOUS APPS OVER THE PAST FOUR YEARS IN TERMS OF AD

| AD Technology | DDF | US | UTP | PPA | ADM | UBS | ADU | UTC |
|---|---|---|---|---|---|---|---|---|
| Quantity | 2,362 | 602 | 387 | 12 | 3 | 1 | 0 | 0 |
| Percentage | 44.6% | 11.4% | 7.3% | 0.2% | 0.1% | 0.02% | 0% | 0% |

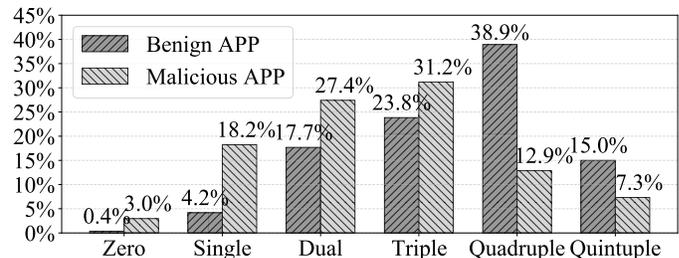

Fig. 8. Different Categories of ARA Techniques Deployed in Benign and Malicious Apps

Table XIII to XV illustrate more detailed aspects. Upon observation, it can be noted that malicious apps exhibit striking similarities to benign apps in terms of the deployment details of various technologies. However, the deployment proportions are generally lower for malicious apps, and they tend to rely on a narrower range of techniques. In the case of AD and VED, some specific technologies have not been detected or identified. The deployment proportion of AH in malicious apps is lower, especially in terms of defense against Frida. This suggests that using modern Hook frameworks for analyzing malicious apps can yield better results. Lastly, in the case of RD, malicious apps tend to lean towards using CSCE, which differs from benign apps. This may be because once a malicious app detects that the current system can execute privileged shell commands, it can perform more malicious actions.

Lastly, we also investigated the preference of benign and malicious apps in utilizing different categories of 5 ARA technologies over the past four years. The specific data is presented in Figure 8, where Zero represents an app that does not employ any ARA technology, while Quintuple indicates an app that utilizes all 5 categories of ARA technologies. The data indicates that the majority of malicious apps tend to utilize 1 to 3 categories of ARA technologies, while benign apps are more inclined to use two or more categories of ARA technologies. Among the benign apps, 53.9% of them employ four or more categories of ARA technologies. Additionally, according to Figure 8, almost all apps deploy at least one category of ARA technology, with a percentage of 99.6% for benign apps and 97.0% for malicious apps.

The analysis findings suggest that the vast majority of malicious apps employ multiple ARA techniques. This presents a greater challenge for security analysts in analyzing malicious apps, as they need to first identify whether the target app utilizes ARA techniques and then gain a deeper understanding of their specific implementation methods. Once the type of ARA technique is determined, security analysts can employ appropriate countermeasures accordingly. For instance, if a certain app refuses to exhibit malicious behavior in a virtual environment, the analyst should choose a real device as the analysis environment or attempt to conceal the characteristics of the virtual environment.

The adoption of multiple ARA techniques by malicious applications poses challenges for security analysts. Nevertheless, with the implementation of targeted countermeasures, it remains possible to analyze and combat these malicious applications effectively.

> **Q3 Finding**
> Except for RD technology, the deployment proportions of the other four categories of technologies are generally lower in malicious apps compared to benign apps. Moreover, compared to benign apps, malicious apps tend to have a higher inclination towards detecting whether the current environment can execute privileged shell commands. Only a few malicious apps have protective measures against Hook frameworks. Lastly, the vast majority of apps utilize at least one category of ARA technology.

## V. DICUSSION

While our study has provided important insights into the deployment of ARA techniques and the characteristics of malicious apps, it is important to acknowledge the limitations of our research.

TABLE XIV
DETAILED ANALYSIS OF MALICIOUS APPS OVER THE PAST FOUR YEARS IN TERMS OF VED

| VED | ED | | | | | | | | | | ALVD | | | |
|---|---|---|---|---|---|---|---|---|---|---|---|---|---|---|
| | ERS | CB | DI | HC | SI | CSBD | MVBD | TM | UESB | TBCBD | FP | SU | CIH | PD |
| Quantity | 1,348 | 685 | 600 | 221 | 75 | 17 | 10 | 0 | 0 | 0 | 135 | 17 | 5 | 3 |
| Percentage | 25.5% | 12.9% | 11.3% | 4.2% | 1.4% | 0.3% | 0.1% | 0% | 0% | 0% | 2.6% | 0.3% | 0.1% | 0.06% |

TABLE XV
DETAILED ANALYSIS OF MALICIOUS APPS OVER THE PAST FOUR YEARS IN TERMS OF AT/AH/RD

| AT/AH/RD | AT | | | | AH | | RD | | | |
|---|---|---|---|---|---|---|---|---|---|---|
| | SC | IV | CIC | SA | Xposed | Frida | CSCE | CIP | CBTSP | DP |
| Quantity | 4,721 | 919 | 291 | 16 | 696 | 7 | 3,390 | 931 | 480 | 212 |
| Percentage | 89.2% | 17.4% | 5.5% | 0.3% | 13.1% | 0.1% | 64.0% | 17.6% | 9.1% | 4.0% |

**Potential Errors.** Despite our rigorous evaluation of the ARA techniques considered in this study and the creation of a ground truth dataset to measure the performance of ARAP, there is still a possibility of false positives or false negatives, which may introduce certain inaccuracies in our findings. Future research could focus on exploring ways to minimize these potential errors and enhance the reliability of our results.

**Dynamic Analyzer to be Strengthened.** In the current implementation of ARAP, we use Monkey to generate random events for app exploration. This approach may result in ARAP not effectively handling some corner cases, such as apps that only perform ARA checks after the user logs into their account. In future research, we aim to refer to solutions for similar issues found in related studies [39], [40] and consider incorporating the latest UI exploration methods [41], [42] to further enhance ARAP's dynamic analysis capabilities.

**Sustained Effectiveness.** In our work, the effectiveness of our tool depends on predefined atomic and fingerprint lists, as mentioned in Section III-A. To capture the latest implementation details of ARA techniques, we conducted a semi-automated static analysis of numerous real-world apps, enabling us to establish an up-to-date atoms and fingerprints list (§III-B). However, given the dynamic nature of ARA technologies, the long-term effectiveness of our tool may be compromised. In our future endeavors, we intend to integrate AI technologies to develop a tool or solution that can autonomously learn and recognize emerging ARA techniques.

## VI. RELATED WORK

In this section, we provide a review of the existing research related to our research questions.

**Partial Summary and Investigation of ARA.** This part of the work focuses solely on certain specific ARA technologies, while the majority of the research either does not cover or only includes a very limited number of real-world apps. This has resulted in a lack of systematic research on ARA technology.

Vikas Sihag et al. [7] presented a comprehensive analysis and classification of available Android malware obfuscation techniques in the literature. Their focus was primarily on obfuscation techniques. They provided theoretical explanations for only a subset of ARA technology and not does it cover specific implementations.

Stefano Berlato et al. [22] conducted a large-scale study on Android applications to quantify the practical application of AD and AT protections. Their work roughly categorizes ARA technology into two classes, providing a general overview of various techniques. Furthermore, their work solely focuses on static methods, limiting their detection capabilities. To illustrate the differences between our work and theirs, we compared ARAP with ATADetector. The results demonstrate that ARAP outperforms ATADetector in terms of overall performance.

**Contradictory dynamic analysis.** This part of the work attempts to study ARA technology using dynamic analysis methods. However, this part of the work relies heavily on Frida, which clearly contradicts common Anti-Hooking methods. This hinders the effectiveness of their research.

Onur Zungur et al. [19] proposed a dynamic analysis framework called AppJITSU to assess the resilience of security-critical applications. They detect ARA technology by observing whether the behavior of an application varies in different environments, but they are unable to accurately determine how the technology is implemented in the application. Additionally, their analysis relies on applications that use Frida [21], which is a popular target for ARA technology. Due to the lack of open-source availability of AppJITSU, further comparisons cannot be made.

Wilco van Beijnum et al. [20] investigated the popularity of obfuscation techniques in mobile applications, which was the first work to combine a study on the popularity of various obfuscation techniques with an analysis across multiple mobile platforms, namely Android and iOS. However, their analysis only provided a superficial examination of each technique, focusing solely on the most representative methods within each category, resulting in an incomplete study. Additionally, their use of Frida to detect anti-hooking techniques presented clear contradictions. As they did not conduct a detailed evaluation of the tool, its effectiveness remains uncertain. Furthermore, the unavailability of HALY hindered further comparisons.

## VII. CONCLUSION

In this paper, we have presented an empirical study on the deployment of ARA techniques for Android apps. To support our research, we have developed a tool called ARAP, which utilizes hybrid analysis to uncover intricate ARA techniques from Android apps. We evaluated the trends and differences in the use of ARA techniques by benign and malicious apps, highlighting the significance of identifying ARA technology in malicious software analysis. Furthermore, we have also provided a dataset of 300 APKs for academic use, which, to the best of our knowledge, is the first available dataset in the relevant field.

In summary, we conduct a large-scale systematic study to identify the current usage of ARA technologies in Android apps. The purpose of our study is to assist security analysts in effectively analyzing malware by bypassing these technologies. By filling this research gap, our findings will contribute to the development of improved Android app protection mechanisms and the detection of malicious Android apps.

## References

[1] IDC, "Smartphone market share," Online, 2020. [Online]. Available: https://www.idc.com/promo/smartphone-market-share/os

[2] G. Kelly. (2014) 97% of mobile malware is on android. this is the easy way you stay safe. [Online]. Available: https://goo.gl/z121Sq

[3] Z. Xie, M. Wen, H. Jia, X. Guo, X. Huang, D. Zou, and H. Jin, "Precise and efficient patch presence test for android applications against code obfuscation," *Proceedings of the 32nd ACM SIGSOFT International Symposium on Software Testing and Analysis*, 2023.

[4] M. Conti, P. Vinod, and A. Vitella, "Obfuscation detection in android applications using deep learning," *J. Inf. Secur. Appl.*, vol. 70, p. 103311, 2022.

[5] R. Guo, Q. Liu, M. Zhang, N. Hu, and H. Lu, "A survey of obfuscation and deobfuscation techniques in android code protection," *2022 7th IEEE International Conference on Data Science in Cyberspace (DSC)*, pp. 40–47, 2022.

[6] M. M. Hammad, J. Garcia, and S. Malek, "A large-scale empirical study on the effects of code obfuscations on android apps and anti-malware products," *2018 IEEE/ACM 40th International Conference on Software Engineering (ICSE)*, pp. 421–431, 2018.

[7] V. K. Sihag, M. Vardhan, and P. Singh, "A survey of android application and malware hardening," *Comput. Sci. Rev.*, vol. 39, p. 100365, 2021.

[8] V. Haupert, D. Maier, N. Schneider, J. Kirsch, and T. Müller, "Honey, i shrunk your app security: The state of android app hardening," in *International Conference on Detection of intrusions and malware, and vulnerability assessment*, 2018.

[9] A. Kalysch, "Android application hardening: Attack surface reduction and ip protection mechanisms," 2020.

[10] D. Jang, Y. Jeong, S. Lee, M. Park, K. Kwak, D. Kim, and B. B. Kang, "Rethinking anti-emulation techniques for large-scale software deployment," *Comput. Secur.*, vol. 83, pp. 182–200, 2019.

[11] K. Tam, A. Feizollah, N. B. Anuar, R. B. Salleh, and L. Cavallaro, "The evolution of android malware and android analysis techniques," *ACM Computing Surveys (CSUR)*, vol. 49, pp. 1 – 41, 2017.

[12] C.-V. Lita, D. Cosovan, and D. Gavrilut, "Anti-emulation trends in modern packers: a survey on the evolution of anti-emulation techniques in upa packers," *Journal of Computer Virology and Hacking Techniques*, vol. 14, pp. 107–126, 2018.

[13] J. Salwan, S. Bardin, and M.-L. Potet, "Symbolic deobfuscation: From virtualized code back to the original," in *Proc. DIMVA*, 2018.

[14] L. Xue, X. Luo, L. Yu, S. Wang, and D. Wu, "Adaptive unpacking of android apps," *2017 IEEE/ACM 39th International Conference on Software Engineering (ICSE)*, pp. 358–369, 2017.

[15] N. S. Evans, A. Benameur, and Y. Shen, "All your root checks are belong to us: The sad state of root detection," in *Proceedings of the 13th ACM International Symposium on Mobility Management and Wireless Access*, 2015, pp. 81–88.

[16] H. Cho, J. Lim, H. Kim, and J. H. Yi, "Anti-debugging scheme for protecting mobile apps on android platform," *The Journal of Supercomputing*, vol. 72, pp. 232–246, 2015.

[17] L. Nguyen-Vu, N.-T. Chau, S. Kang, and S. Jung, "Android rooting: An arms race between evasion and detection," *Secur. Commun. Networks*, vol. 2017, pp. 4 121 765:1–4 121 765:13, 2017.

[18] L. Shi, J. Fu, Z. Guo, and J. Ming, ""jekyll and hyde" is risky: Shared-everything threat mitigation in dual-instance apps," *Proceedings of the 17th Annual International Conference on Mobile Systems, Applications, and Services*, 2019.

[19] O. Zungur, A. Bianchi, G. Stringhini, and M. Egele, "Appjitsu: Investigating the resiliency of android applications," *2021 IEEE European Symposium on Security and Privacy (EuroS&P)*, pp. 457–471, 2021. [Online]. Available: https://api.semanticscholar.org/CorpusID:232361273

[20] A. Beijnum, "Haly: Automated evaluation of hardening techniques in android and ios apps," Master's thesis, University of Twente, 2023.

[21] "Frida," [Online] https://github.com/frida/frida.

[22] S. Berlato and M. Ceccato, "A large-scale study on the adoption of anti-debugging and anti-tampering protections in android apps," *J. Inf. Secur. Appl.*, vol. 52, p. 102463, 2020.

[23] L. Xue, H. Zhou, X. Luo, L. Yu, D. Wu, Y. Zhou, and X. Ma, "Packergrind: An adaptive unpacking system for android apps," *IEEE Transactions on Software Engineering*, vol. 48, pp. 551–570, 2020.

[24] Z. Dong, H. Liu, L. Wang, X. Luo, Y. Guo, X. Xu, X. Xiao, and H. Wang, "What did you pack in my app? a systematic analysis of commercial android packers," *Proceedings of the 30th ACM Joint European Software Engineering Conference and Symposium on the Foundations of Software Engineering*, 2022.

[25] T. K. Apostolopoulos, V. Katos, K. R. Choo, and C. Patsakis, "Resurrecting anti-virtualization and anti-debugging: Unhooking your hooks," *Future Gener. Comput. Syst.*, vol. 116, pp. 393–405, 2021.

[26] J. Wan, M. Zulkernine, and C. Liem, "A dynamic app anti-debugging approach on android art runtime," *2018 IEEE 16th Intl Conf on Dependable, Autonomic and Secure Computing, 16th Intl Conf on Pervasive Intelligence and Computing, 4th Intl Conf on Big Data Intelligence and Computing and Cyber Science and Technology Congress(DASC/PiCom/DataCom/CyberSciTech)*, pp. 560–567, 2018.

[27] A. Guerra-Manzanares, H. Bahsi, and S. Nõmm, "Differences in android behavior between real device and emulator: A malware detection perspective," *2019 Sixth International Conference on Internet of Things: Systems, Management and Security (IOTSMS)*, pp. 399–404, 2019.

[28] O. Sahin, A. K. Coskun, and M. Egele, "Proteus: Detecting android emulators from instruction-level profiles," in *International Symposium on Recent Advances in Intrusion Detection*, 2018.

[29] X. Wang, S. Zhu, D. Zhou, and Y. Yang, "Droid-antirm: Taming control flow anti-analysis to support automated dynamic analysis of android malware," *Proceedings of the 33rd Annual Computer Security Applications Conference*, 2017.

[30] V. M. Afonso, A. Kalysch, T. Müller, D. Oliveira, A. R. A. Grégio, and P. L. de Geus, "Lumus: Dynamically uncovering evasive android applications," in *Information Security Conference*, 2018.

[31] F. Xu, S. Shen, W. Diao, Z. Li, Y. Chen, R. Li, and K. Zhang, "Android on pc: On the security of end-user android emulators," *Proceedings of the 2021 ACM SIGSAC Conference on Computer and Communications Security*, 2021.

[32] Y. Hong, Y. Hu, C.-M. Lai, S. F. Wu, I. Neamtiu, P. Mcdaniel, P. L. Yu, H. Çam, and G.-J. Ahn, "Defining and detecting environment discrimination in android apps," in *Security and Privacy in Communication Networks*, 2017.

[33] D. Dai, R. Li, J. Tang, A. Davanian, and H. Yin, "Parallel space traveling: A security analysis of app-level virtualization in android," *Proceedings of the 25th ACM Symposium on Access Control Models and Technologies*, 2020.

[34] C. Zheng, T. Luo, Z. Xu, W. Hu, and X. Ouyang, "Android plugin becomes a catastrophe to android ecosystem," *Proceedings of the First Workshop on Radical and Experiential Security*, 2018.

[35] L. Li, T. F. Bissyandé, and J. Klein, "Rebooting research on detecting repackaged android apps: Literature review and benchmark," *IEEE Transactions on Software Engineering*, vol. 47, pp. 676–693, 2018.

[36] A. Ruggia, E. Losiouk, L. Verderame, M. Conti, and A. Merlo, "Repack me if you can: An anti-repackaging solution based on android virtualization," *Annual Computer Security Applications Conference*, 2021.

[37] N. Totosis and C. Patsakis, "Android hooking revisited," *2018 IEEE 16th Intl Conf on Dependable, Autonomic and Secure Computing, 16th Intl Conf on Pervasive Intelligence and Computing, 4th Intl Conf on Big Data Intelligence and Computing and Cyber Science and Technology Congress(DASC/PiCom/DataCom/CyberSciTech)*, pp. 552–559, 2018.

[38] "Androguard," Online, 2016. [Online]. Available: https://github.com/androguard/androguard

[39] F. He, Y. Jia, J. Zhao, Y. Fang, J. Wang, M. Feng, P. Liu, and Y. Zhang, "Maginot line: Assessing a new cross-app threat to pii-as-factor authentication in chinese mobile apps," *Proceedings 2024 Network and Distributed System Security Symposium*, 2024. [Online]. Available: https://api.semanticscholar.org/CorpusID:267624674

[40] S. Höltervennhoff, N. Wöhler, A. Möhle, M. Oltrogge, Y. Acar, O. Wiese, and S. Fahl, "A mixed-methods study on user experiences and challenges of recovery codes for an end-to-end encrypted service," in *USENIX Security Symposium*, 2024. [Online]. Available: https://api.semanticscholar.org/CorpusID:268524085

[41] X. Zhang and et al., "Scene-driven exploration and gui modeling for android apps," in *Proceedings of the 38th IEEE/ACM International Conference on Automated Software Engineering (ASE)*. IEEE, 2023.

[42] M. Auer and G. Fraser, "Exploring android apps using motif actions," in *2023 38th IEEE/ACM International Conference on Automated Software Engineering Workshops (ASEW)*. IEEE, 2023, pp. 135–142.

# APPENDIX A
# ANTI-DEBUGGING (AD)

Anti-debugging is a technique used to protect applications from malicious analysts' attacks and reverse engineering. This technique involves inserting anti-debugging code into the application code, making it difficult for malicious analysts to use debuggers and other tools to debug and reverse engineer the application. Deploying appropriate anti-debugging techniques can effectively enhance the security and resilience of applications against malicious attacks.

## A. Using Timing Check (UTC)

The Using Timing Check technique detects the presence of a debugger by leveraging the speed difference between the execution of code by the debugger and the debugged application. Developers insert a code segment within the application that performs a calculation and returns a result. Under normal circumstances, this code segment should complete the calculation and return the result in a very short time. However, if the application is being debugged, the debugger may interfere with the execution of the code, causing the calculation to take longer.

## B. Using Signals (US)

The Using Signals technique exploits the difference in signal handling between the debugged application and the debugger. Developers insert a code segment within the application that sends a SIGTRAP signal to itself. Under normal circumstances, the application captures and handles this signal, but if the application is being debugged, the debugger captures the signal and pauses the application's execution to facilitate debugging. Therefore, the Using Signals technique detects whether an application is being debugged by checking if it has received a SIGTRAP signal. Other signals with similar functionality to SIGTRAP can also accomplish this task.

## C. Using TracerPID (UTP)

The TracerPID status value of an application is recorded in the /proc/PID/status file. Under normal circumstances, the TracerPID status value should be 0, indicating that the application is not being traced. However, if the application is being debugged, the TracerPID status value of the application changes to the PID value of the debugger process attached to it. Therefore, by checking whether the TracerPID status value in the /proc/PID/status file is zero, the application can determine whether it is being debugged.

## D. Preempt Ptrace API (PPA)

In the Android system, each process has a unique Process ID (PID). When a process calls the ptrace system call and passes its own PID as a parameter, it is marked as being debugged. Since a process can only be attached by one process at a time, if an application has attached itself, no other process can attach to that application, not even a debugger. Therefore, developers can proactively attach themselves to the application at startup to prevent a debugger from attaching to the program.

## E. Anti-Dumping (ADU)

Anti-dumping is used to prevent an application's code from being dumped to disk, thus preventing attackers from analyzing the code. Inotify event monitoring of dumping is a common implementation method. Inotify is a file system event notification mechanism provided by the Linux kernel that can monitor changes to files or directories and notify the corresponding process when a change occurs. Developers can create an Inotify instance using the Inotify API and monitor the target file or directory by calling the *Inotify_add_watch* function. Developers can determine whether a process is attempting to dump the application's code by capturing Inotify events and can take appropriate measures, such as self-destruction or triggering an alert.

## F. Debuggable and Debugger Features (DDF)

The Android system offers a range of APIs to detect the presence of a debugger. These APIs, such as isDebuggerConnected and waitingForDebugger, allow for quick determination of whether a debugger is connected. Furthermore, the existence of a debugger can be identified by examining specific indicators, such as checking for the utilization of port 23946 or searching for processes named "android_server" or "gdb_server," among others.

## G. Using Breakpointset Structure (UBS)

The Breakpointset Structure is used to store information about all breakpoints set in a process, including their addresses, number, and other detailed information. Developers can use this structure to detect and block the debugger from working. For example, developers can check the number of breakpoints set in the program, and if it is not zero, assume that the program may be in debug mode. Alternatively, they can intentionally assign a null pointer to the breakpoint set, causing the process to terminate due to incorrect memory access, thus preventing the debugger from running.

## H. Altering Debugger Memory (ADM)

Altering Debugger Memory is a method of manipulating the behavior of a debugger by modifying the global virtual machine state of a running application. In Dalvik, this could be achieved by accessing the DvmGlobals structure pointed to by the global variable gDvm. In ART, although the gDvm variable is no longer available, the ART runtime exports some JDWP-related pointers as global symbols. Applications can modify the behavior of the debugger by overwriting these variables, such as replacing the function pointer that handles debugger with a pointer to a function that always returns false, or replacing the address of the function *jdwpAdbState::ProcessIncoming* with the address of the function *jdwpAdbState::Shutdown*, causing the debugger to immediately

# APPENDIX B
# VIRTUAL ENVIRONMENT DETECTION (VED)

The main purpose of this technique is to detect whether a device is running in a virtual environment. For malicious applications, by detecting whether the device is running in a virtual environment, they can decide whether to exhibit malicious behavior to evade detection. For benign applications, developers can determine whether their application is running in a secure environment. When a virtual environment is detected, developers can take some countermeasures according to their needs, such as denying the application to run in a virtual environment to protect the security and integrity of the application.

*A. Emulator Detection (ED)*

Emulator Detection refers to a type of technology that is used to detect and defend against applications running in an emulator or virtual machine environment, to prevent malicious attacks, anti-debugging, cracking, or other fraudulent behavior. Emulator Detection technology typically detects features of the emulator or virtual machine and takes corresponding actions when these features are detected, such as shutting down the application or displaying warning messages.

*1) Device Identifier (DI):* In the Android system, each device has a unique device identifier called Android ID. However, in emulators or virtual machines, the same Android ID is often used, so it is possible to detect whether the current device is an emulator by checking if the current Android ID matches the commonly used Android ID in emulators.

*2) Current Build (CB):* Detecting the current build is one of the common anti-emulator techniques. The current build refers to the current operating system version and its related parameters, such as kernel version, firmware version, compilation time, etc. In emulators or virtual machines, specific build parameters are often used, so detecting whether the current build matches the current build of a real device can determine whether the current device is an emulator.

*3) Telephony Manager (TM):* Telephony Manager is a system service in the Android operating system that provides information related to mobile communication, such as the device's IMSI, IMEI, phone number, etc. In emulators or virtual machines, real mobile communication functionality is usually not provided, so it is possible to detect whether the current device is an emulator by checking whether the information provided by Telephony Manager is empty or matches the information of a real device.

*4) Hardware Components (HC):* When an application runs in an emulator or virtual machine environment, it usually simulates some hardware components, and the information of these components may be different from that of a real device. Therefore, it is possible to determine whether the current device is an emulator by reading the hardware information provided by the emulator. These hardware information can be obtained by reading the hardware information files (such as /proc/cpuinfo, /proc/meminfo, /proc/version, etc.) in the /proc file system and the CPU-related files (such as cpuinfo_max_freq, cpuinfo_min_freq, etc.) under the /sys/devices/system/CPU/cpu0/cpufreq/ directory. In addition to CPU information, other hardware component information can also be obtained, such as device screen resolution, RAM size, cameras, etc. If the information of these hardware components does not match the real device, it can be inferred that the current device may be an emulator or virtual machine.

*5) Emulator Related Strings (ERS):* This technique leverages the differences between emulators and real devices in certain strings and system properties to detect whether the current device is an emulator. Some common features include strings such as "generic" and "Genymotion", system properties such as "ro.kernel.qemu", and files such as "/dev/socket/qemud". These strings, properties, and files are often present in emulator environments but rarely or never found on real devices.

*6) Sensor Information (SI):* In a simulator, sensor data is often fixed or random due to the lack of real sensor hardware. On a real device, sensor data typically varies based on the device's actual movement and changes in the surrounding environment. Therefore, by obtaining and analyzing sensor data, an application can determine if the current device is a simulator. Some commonly used sensors include accelerometers, gyroscopes, light sensors, and so on.

*7) CPU Performance:* When running applications on a simulator, the simulator needs to try to simulate CPU execution on a real device. However, due to hardware and software differences between the simulator and the real device, as well as performance limitations of the simulator, it is difficult for the simulator to completely simulate CPU execution on a real device, which may result in execution differences. QEMU is a commonly used open-source emulator that can simulate various computer architectures. In Android development, QEMU is also widely used as an Android emulator. It can simulate Android device components such as the CPU and memory, and can run on different platforms. Due to its flexibility and customizability, QEMU is widely used in Android development and security. In this paper, we only consider QEMU-based emulators.

• *Context Switch Based Detection (CSBD).* In QEMU, instruction translation is cached into a data structure called "translation block cache" to accelerate the execution of the emulator in subsequent runs. There will be no race conditions within the same basic block, even without locking. To detect the presence of an emulator, a code segment that triggers a race condition can be constructed for testing. On a real device, this code will cause a context switch and thus a race condition. However, in the emulator, no race condition will be observed.

• *Translation Block Cache Based Detection (TBCBD).* Translation Block Cache Based Detection utilizes QEMU's Translation Block (TB) caching mechanism. When QEMU runs, the instructions it executes are translated into a series of basic blocks, which are cached in the TB for improved performance. When the same instructions are executed again, QEMU will call the cached basic blocks from the TB, without needing to decode and translate the instructions again. When executing self-modifying code in the emulator, these codes will break the TB cache because the cached basic blocks may no longer be valid. This will cause QEMU to have to decode and translate instructions again, resulting in a significant increase in execution time. On a real device, since there is no TB caching mechanism, executing self-modifying code will not cause similar performance degradation.

- *Misaligned Vectorization Based Detection (MVBD)*. On real devices, certain high-performance vectorized instructions like Intel SIMD and ARM NEON do not support unaligned memory access, which can lead to application errors. However, in a simulator, since memory access is ultimately reconstructed at the software level, unaligned vectorized operations can be executed normally without causing application errors, allowing this difference to be used to distinguish between real devices and simulators.

- *Updates to Execution State Bits (UESB)*. The CPSR (Current Program Status Register) is a special register in the ARM architecture that records the current execution status of the processor, including program status, control status, and condition status, among others. When updating the CPSR register using the MSR system instructions, updating other execution status bits is prohibited, but updating the CPSR.E (i.e., byte order flag) is allowed. However, QEMU ignores the MSR instructions' update of the CPSR.E bit, thus identifying the emulator by attempting to flip the CPSR.E bit and checking if the byte order has been successfully changed.

### B. App-Level Virtualization Detection (ALVD)

Android App-Level Virtualization technology is an innovative virtualization technique that allows loading and running any client APK file within a host application, improving the flexibility and interoperability of applications. It takes over the system services by intercepting them to achieve a more flexible runtime environment. Unlike traditional emulators that typically run virtual machines on computers to simulate device environments, App-Level Virtualization technology creates a virtual environment on a real device to run applications. Therefore, some traditional emulator detection methods may fail to detect it.

*1) File Path (FP):* Detecting the use of App-Level Virtualization is commonly done by checking file paths. One method is to check if the current application's path is included in the paths of multi-instance applications. Multi-instance applications usually store the files of their included applications in different directories. Therefore, if the current application's path is included in the paths of multi-instance applications, it can be assumed that the current application is running in a virtualized environment. Another method is to search for two different "base.apk" paths in the process memory: one belonging to the guest application and the other to the host application. If two different paths are found, it can be assumed that the current application is running in a virtualized environment.

*2) Share UID (SU):* When using App-Level Virtualization technology, the host and guest applications usually share the same User ID (UID). This is because the host application is responsible for providing the virtualization environment for the guest applications and must therefore share the same system permissions as the guest applications. Therefore, by checking whether there are multiple applications on a device with the same UID, it is possible to detect whether App-Level Virtualization technology is being used.

*3) Port Detection (PD):* In this technique, the application will always listen to a specific port at runtime and send a message to that port during initialization. Once the application receives this message, it will automatically close. If attempting to start another instance, the same process will be repeated, and the previous instance will be closed on the new instance. This achieves the effect of preventing the application from being opened multiple times.

*4) Code Injection and Hooking (CIH):* Code Injection and Hooking techniques can be used to detect App-Level Virtualization, with two methods available. The first method involves directly searching for suspicious shared object (so) libraries, the host application usually employs its own so libraries to manage guest applications. Hence, searching for suspicious strings or functions in an application's so library is a commonly used detection method. The second method involves using stack tracing to determine if the callActivityOnCreate method is executed multiple times. In App-Level Virtualization, the host application rewrites the start function of an Activity by hooking system calls, thus achieving the capability to launch multiple instances of an Activity. As such, detecting the presence of App-Level Virtualization can be achieved by checking whether the *callActivityOnCreate* method is executed multiple times in the stack.

## APPENDIX C
## ANTI-TAMPERING (AT)

"Anti-Tampering" is designed to prevent malicious attackers from tampering or repackaging applications. Implementing an Anti-Tampering mechanism in an application can effectively protect it from malicious modifications, thereby ensuring the security and integrity of the application.

### A. Signature Checking (SC)

The working principle of Signature Checking is to verify the integrity and authenticity of an application by examining its digital signature. In the Android system, an application's digital signature is generated by the application developer using a private key, which serves as proof that the application comes from the developer and has not been tampered with. Typically, an application's digital signature in the Android system is stored in the META-INF directory of the APK file, in a file named CERT.RSA. Specifically, the Android V2 and later signature mechanisms have strong anti-tampering capabilities, which can effectively ensure the integrity of the APK. In addition, developers can further compare the current app's signature with the one stored on a remote server during the program's runtime to ensure the integrity of the APK.

### B. Code Integrity Checking (CIC)

Code Integrity Checking refers to the process of verifying the integrity of an application's code during runtime. Developers can actively calculate the hash value of a specific file during runtime and compare it with the expected value. If the two values differ, it indicates that the application may have been tampered with. At this point, developers can choose to prompt the user with relevant error information and close the application.

*C. SafetyNet Attestation (SA)*

SafetyNet Attestation is a security validation service for Android applications introduced by Google, aimed at helping developers protect their applications from malicious attacks and tampering. It can check the integrity and authenticity of applications, ensuring they have not been tampered with or maliciously modified. Developers can add a dependency on SafetyNet Attestation and use the SafetyNet API in their applications. It is worth noting that Google will gradually discontinue the use of the SafetyNet Attestation API starting in 2024 and encourage the use of the Play Integrity API, which is a more secure solution. We also consider the Play Integrity API in our work.

*D. Installer Verification (IV)*

The purpose of Installer Verification is to ensure that the installation program is from a trusted source. For example, if a developer only uploads an application to the Google Play Store, but the installation source is from another app market, it may indicate a risk of repackaging the application. Developers can enable Installer Verification by adding "android:verification="true"" to the application manifest file (AndroidManifest.xml).

APPENDIX D
ANTI-HOOKING (AH)

Android Hooking techniques refer to the use of hooking technology by attackers to monitor, modify, or tamper with the behavior of an application, thereby carrying out malicious attacks or stealing sensitive information. Commonly used tools include Xposed and Frida.

*A. Xposed*

Xposed is an open-source hooking framework that allows developers to modify the behavior of applications by writing modules. Common techniques to detect the Xposed environment include checking for the existence of the XposedBridge.jar file in the system, verifying the installation of the Xposed Installer application, and checking for the presence of the XposedHelpers class.

*B. Frida*

Frida is a dynamic instrumentation tool that allows for hooking into applications using JavaScript scripts. It can be used for monitoring and modifying aspects of an application such as memory, files, and network activity. To detect the presence of a Frida environment, the existence of the fridaserver binary file, the LIBFRIDA library file, and the frida-agent file can be checked. The fridaserver and LIBFRIDA library files are the core components of Frida, while frida-agent is a proxy tool for Frida. The presence of these files indicates that a Frida environment is installed or that the proxy has been started.

APPENDIX E
ROOT DETECTION (RD)

In the Android system, Root refers to the process of gaining superuser privileges. With Root, users can access all the features and files of the system, and thus can modify system settings, delete pre-installed applications, install unverified applications, and so on.

*A. Check Installed Packages (CIP)*

Check Installed Packages is a root detection method that determines if a device has been rooted by checking if known root tools or files are installed on the device, such as the su binary file, SuperSU or SuperUser app, Magisk, and more. This method utilizes the PackageManager class in Android system to obtain a list of installed applications on the device, and performs comparison and analysis to detect the root status.

*B. Check Shell Commands Execution (CSCE)*

Check Shell Commands Execution is a common Root detection method that detects whether a device has been rooted by executing some Linux commands that require Root privileges. In the Android system, some Linux commands such as "su" and "busybox" can only run after the device has been rooted. Therefore, by attempting to execute these Root-required commands and observing the returned results, it is possible to infer whether the device has been rooted.

*C. Check Build Tag and System Properties (CBTSP)*

The Check Build Tag and System Properties is a Root detection method that determines whether a device has been rooted by examining the device's "Build Tag" and system properties. In the Android system, the "Build Tag" is a string used to identify the firmware version of the device. System properties, on the other hand, are a set of key-value pairs used to store some properties and configuration information of the Android system. These properties and firmware version information may be modified on a rooted device.

*D. Directory Permissions (DP)*

Directory Permissions is a root detection method that determines if a device has been rooted by checking the permissions of certain directories in the Android file system. After a device is rooted, the permissions of certain directories may be changed to allow normal users to read, write, and execute operations on them.